\documentclass[twoside]{ilcws07}
\usepackage[latin1]{inputenc}
\usepackage[dvips]{graphicx,epsfig,color}
\usepackage{wrapfig,rotating}
\usepackage{amssymb,amsmath,array,cite,psfrag}

\input{paperdef}

\begin{document}


\title{
$\cp$-violating Loop Effects in the\\
Higgs Sector of the MSSM}
\author{T.~Hahn$^1$, S.~Heinemeyer$^2$, W.~Hollik$^1$,
H.~Rzehak$^3$, G.~Weiglein$^4$ and K.E.~Williams$^4$
\vspace{.3cm}\\
1- Max-Planck-Institut f\"ur Physik, 
F\"ohringer Ring 6, D--80805 Munich, Germany \\[.1cm]
2- Instituto de Fisica de Cantabria (CSIC-UC), Santander,  Spain \\[.1cm]
3- Paul Scherrer Institut, W\"urenlingen und Villigen, CH--5232
Villigen PSI, Switzerland\\[.1cm]
4- IPPP, University of Durham, Durham DH1~3LE, UK
}

\maketitle

\begin{abstract}
$\cp$-violating effects in the Higgs sector of the Minimal
Supersymmetric Standard Model with complex parameters (cMSSM) are induced 
by potentially large higher-order corrections. As a consequence, all 
three neutral Higgs bosons can mix with each other. 
Recent results for loop corrections
in the Higgs sector of the cMSSM are reviewed~\cite{url}. 
Results for 
propagator-type corrections of \order{\alt\als}
and complete one-loop
results for Higgs cascade decays of the kind $h_a \to h_b h_c$
are summarised, and the proper treatment of external Higgs
bosons in Higgs-boson production
and decay processes is discussed.
\end{abstract}

\section{Introduction}

A striking prediction of models of 
supersymmetry (SUSY) is a Higgs sector with at
least one relatively light Higgs boson. In the Minimal Supersymmetric
extension of the Standard Model (MSSM) two Higgs doublets are required,
resulting in five physical Higgs bosons. 
In lowest order these are the light and heavy $\cp$-even $h$
and $H$, the $\cp$-odd $A$, and the charged Higgs bosons $H^\pm$.
The Higgs sector of the MSSM can be characterised at lowest
order by the two parameters (besides the gauge couplings)
$\MHp$ and $\tb \equiv v_2/v_1$, 
the ratio of the two vacuum expectation values. All other masses and
mixing angles can be predicted in terms of these parameters. 
Higher-order contributions yield large corrections to the tree-level 
relations and, via complex phases, induce $\cp$-violating effects. 
In the MSSM with complex parameters (cMSSM) therefore all three neutral
Higgs bosons can mix with each other. The corresponding mass eigenstates
are denoted as
$h_1$, $h_2$, $h_3$. If the mixing between the three neutral mass
eigenstates is such that the coupling of the lightest Higgs boson to
gauge bosons is significantly suppressed, this state can be very light
without being in conflict with the exclusion bounds from the LEP Higgs
searches~\cite{LEPHiggsSM,LEPHiggsMSSM}. In this case the second-lightest 
Higgs boson,
$h_2$, may predominantly decay into a pair of light Higgs bosons,
$h_2 \to h_1h_1$.

We report in this paper on recent progress on higher-order corrections
in the Higgs sector of the cMSSM%
\footnote{See e.g.\ \citeres{PomssmRep,mhiggsWN,abdel,sven} 
for recent reviews of
the present status of higher-order corrections in the Higgs sector of
the MSSM with and without complex phases.}. We briefly discuss
propagator-type corrections of \order{\alt\als}~\cite{mhcMSSM2L}
and complete one-loop
results for Higgs cascade decays of the kind $h_a \to h_b h_c$
($a,b,c = 1,2,3$)~\cite{hdeccpv}. In this context we put a particular
emphasis on the treatment of external Higgs states
in Higgs-boson production and decay process in the presence of
$\cp$-violating mixing among all three neutral Higgs bosons.


\section{External on-shell Higgs-bosons}

The propagator matrix of the neutral Higgs bosons $h, H, A$ can be
written as a $3 \times 3$ matrix, $\De_{hHA}(p^2)$
(we neglect mixing with the Goldstone boson
$G$ and the $Z$~boson in the propagator matrix since
the corresponding contributions are of sub-leading two-loop order,
see the discussion in \citere{mhcMSSMlong}).
This propagator matrix
is related to the $3 \times 3$ matrix of the irreducible
vertex functions by
\begin{equation}
\De_{hHA}(p^2) = - \left(\hat{\Gamma}_{hHA}(p^2)\right)^{-1} ,
\label{eq:propagator}
\end{equation}
where $\hat{\Gamma}_{hHA}(p^2) = i \left[p^2 \unity -
\matr{M}_{\mathrm{n}}(p^2) \right]$,
\begin{align}
  \matr{M}_{\mathrm{n}}(p^2) &=
  \begin{pmatrix}
    \mh^2 - \ser{hh}(p^2) & - \ser{hH}(p^2) & - \ser{hA}(p^2) \\
    - \ser{hH}(p^2) & \mH^2 - \ser{HH}(p^2) & - \ser{HA}(p^2) \\
    - \ser{hA}(p^2) & - \ser{HA}(p^2) & \mA^2 - \ser{AA}(p^2)
  \end{pmatrix}. 
\label{eq:Mn}
\end{align}
Here $m_i$ ($i = h,H,A$) denote the tree-level Higgs-boson masses, and
$\ser{ij}$ are the renormalised self-energies. 
Inversion of $\hat{\Gamma}_{hHA}(p^2)$ yields for the diagonal Higgs
propagators ($i = h, H, A$)
\begin{equation}
\De_{ii}(p^2) = \frac{i}{p^2 - m_i^2 + \ser{ii}^{\rm eff}(p^2)} ,
\label{eq:higgsprop}
\end{equation}
where $\De_{hh}(p^2)$, $\De_{HH}(p^2)$, $\De_{AA}(p^2)$ are the $(11)$,
$(22)$, $(33)$ elements of the $3 \times 3$ matrix $\De_{hHA}(p^2)$,
respectively. The structure of \refeq{eq:higgsprop} is formally the same
as for the case without mixing, but the usual self-energy is replaced by
the effective quantity $\ser{ii}^{\rm eff}(p^2)$ which contains mixing
contributions of the three Higgs bosons. It reads (no summation over 
$i, j, k$)
\begin{align}
\ser{ii}^{\rm eff}(p^2) &= \ser{ii}(p^2) - i 
\frac{2 \hat{\Gamma}_{ij}(p^2) \hat{\Gamma}_{jk}(p^2) \hat{\Gamma}_{ki}(p^2) -
      \hat{\Gamma}^2_{ki}(p^2) \hat{\Gamma}_{jj}(p^2) -
      \hat{\Gamma}^2_{ij}(p^2) \hat{\Gamma}_{kk}(p^2)
     }{\hat{\Gamma}_{jj}(p^2) \hat{\Gamma}_{kk}(p^2) - 
       \hat{\Gamma}^2_{jk}(p^2)
      } ,
\label{eq:sigmaeff}
\end{align}
where the $\hat{\Gamma}_{ij}(p^2)$ are the elements of the $3 \times 3$
matrix $\hat{\Gamma}_{hHA}(p^2)$ as specified above.
The expressions for the off-diagonal 
Higgs propagators read ($i, j, k$ all different, no summation over 
$i, j, k$)
\begin{align}
\De_{ij}(p^2) = \frac{\hat{\Gamma}_{ij} \hat{\Gamma}_{kk} -
                      \hat{\Gamma}_{jk} \hat{\Gamma}_{ki}}{
   \hat{\Gamma}_{ii}\hat{\Gamma}_{jj}\hat{\Gamma}_{kk}
   + 2 \hat{\Gamma}_{ij}\hat{\Gamma}_{jk}\hat{\Gamma}_{ki}
   - \hat{\Gamma}_{ii}\hat{\Gamma}_{jk}^2
   - \hat{\Gamma}_{jj}\hat{\Gamma}_{ki}^2
   - \hat{\Gamma}_{kk}\hat{\Gamma}_{ij}^2
                     } ,
\label{eq:higgsprop2}
\end{align}
where we have dropped the argument $p^2$ of the $\hat{\Gamma}_{ij}(p^2)$
appearing on the right-hand side for ease of notation.
The three complex poles ${\cal M}^2$ of $\De_{hHA}$, \refeq{eq:propagator},
are defined as the solutions of
\begin{equation}
{\cal M}_i^2 - m_i^2 + \ser{ii}^{\rm eff}({\cal M}_i^2) = 0,
~i = h,H,A ,
\label{eq:massmaster}
\end{equation}
with a decomposition of the complex pole as
${\cal M}^2 = M^2 - i M \Ga$,
where $M$ is the mass of the particle and $\Ga$ its width.
We define the loop-corrected mass eigenvalues according to
$M_{h_1} \leq M_{h_2} \leq M_{h_3}$.

We now turn to the on-shell properties of an in- or out-going Higgs boson.
In order to ensure the correct on-shell properties of S-matrix elements
involving external Higgs it is convenient to introduce finite wave
function normalisation factors $\hat Z_i$, $\hat Z_{ij}$
(``Z-factors'').
A vertex with an external Higgs boson, $i$,
can be written as (with $i,j,k$ all different, $i,j,k = h, H, A$, and no
summation over indices)
\BE
\sqrt{\hat Z_i} \KL \Ga_i \; + \; 
          \hat Z_{ij} \Ga_j \; + \; \hat Z_{ik} \Ga_k + \ldots \KR ~,
\label{eq:zfactors}
\end{equation}
where the ellipsis represents contributions from the mixing with the
Goldstone boson and the $Z$~boson, see \citeres{mhcMSSMlong,hdeccpv}.
The Z-factors are given by: 
\begin{equation}
\hat Z_i = \frac{1}{1 + 
              \left(\ser{ii}^{\rm eff}\right)^\prime(\cM_i^2)} , 
\quad
%
\hat Z_{ij} = \frac{\De_{ij}(p^2)}{\De_{ii}(p^2)}_{~\Bigr| p^2 = \cM_i^2} 
               \label{eq:zizij}
\end{equation}
where the propagators $\De_{ii}(p^2)$, $\De_{ij}(p^2)$ 
have been given in \refeqs{eq:higgsprop} and (\ref{eq:higgsprop2}),
respectively. 
The Z-factors can be expressed in terms of a (non-unitary) matrix
$\matr{\hat Z}$, whose elements take the form
(with $\hat Z_{ii} = 1$, $i, j = h, H, A$, and no summation over $i$)
\begin{equation}
(\matr{\hat Z})_{ij} := \sqrt{\hat Z_i} \; \hat Z_{ij}~.
\label{eq:defZ}
\end{equation}
A vertex with one external Higgs boson $h_1$, for instance, is then given by
\BE
(\matr{\hat Z})_{hh} \Ga_h +
(\matr{\hat Z})_{hH} \Ga_H +
(\matr{\hat Z})_{hA} \Ga_A + \ldots ~,
\label{eq:zfactors123}
\end{equation}
where the ellipsis again represents contributions from the mixing with the
Goldstone boson and the $Z$~boson.

It should be noted that the definition of the Z-factors
used here and in \citere{hdeccpv}
differs slightly from the one in \citere{mhcMSSMlong}.
The Higgs-boson self-energies in \refeq{eq:zizij} are
evaluated at the
complex pole, whereas in \citere{mhcMSSMlong} the real part of the 
complex pole had been used. Furthermore, in the definition of $\hat Z_i$
in \refeq{eq:zizij}  $\ser{ii}^{\rm eff}$ appears, as compared to 
$\re\ser{ii}^{\rm eff}$ in \citere{mhcMSSMlong}. While the contributions of
the imaginary parts in \refeq{eq:zizij} to Higgs-boson
production and decay processes are formally of sub-leading two-loop
order, it turns out that their inclusion in general improves the
numerical stability of the results.


\section{Propagator-type corrections of \boldmath{\order{\alt\als}}}

\psfrag{Mh1 [GeV]}{ $M_{h_1}$ [GeV]}
\psfrag{phiAt}{$\phiat/\pi$}
\psfrag{phiXt}{$\varphi_{X_t}/\pi$}
\psfrag{|At| = 1600 GeV}{\hspace*{-0.3cm}\small $|\At| = 1.6 \tev$}
\psfrag{At = 2600 GeV}{\hspace*{-0.1cm}\small $|\At| = 2.6 \tev$}
\psfrag{|Xt| = 1500 GeV}{\hspace*{-0.3cm}\small $|X_t| = 1.5 \tev$}
\psfrag{Xt = 2500 GeV}{\hspace*{-0.3cm}\small $|X_t| = 2.5 \tev$}
\psfrag{MHp = 500 GeV}{\hspace*{-0.1cm}\small $M_{H^\pm} = 500 \gev$}
\psfrag{1 loop}{\raisebox{0.5ex}{\hspace*{-0.5cm} $\mathcal O(\al)$}}
\psfrag{1l+ 2 loop}{\raisebox{0.2ex}{\hspace*{-0.8cm}
    $\mathcal O(\al +\alt \als)$}}
\begin{figure}[htb!]
\centerline{
\includegraphics[width=0.58\textwidth,height=14em]{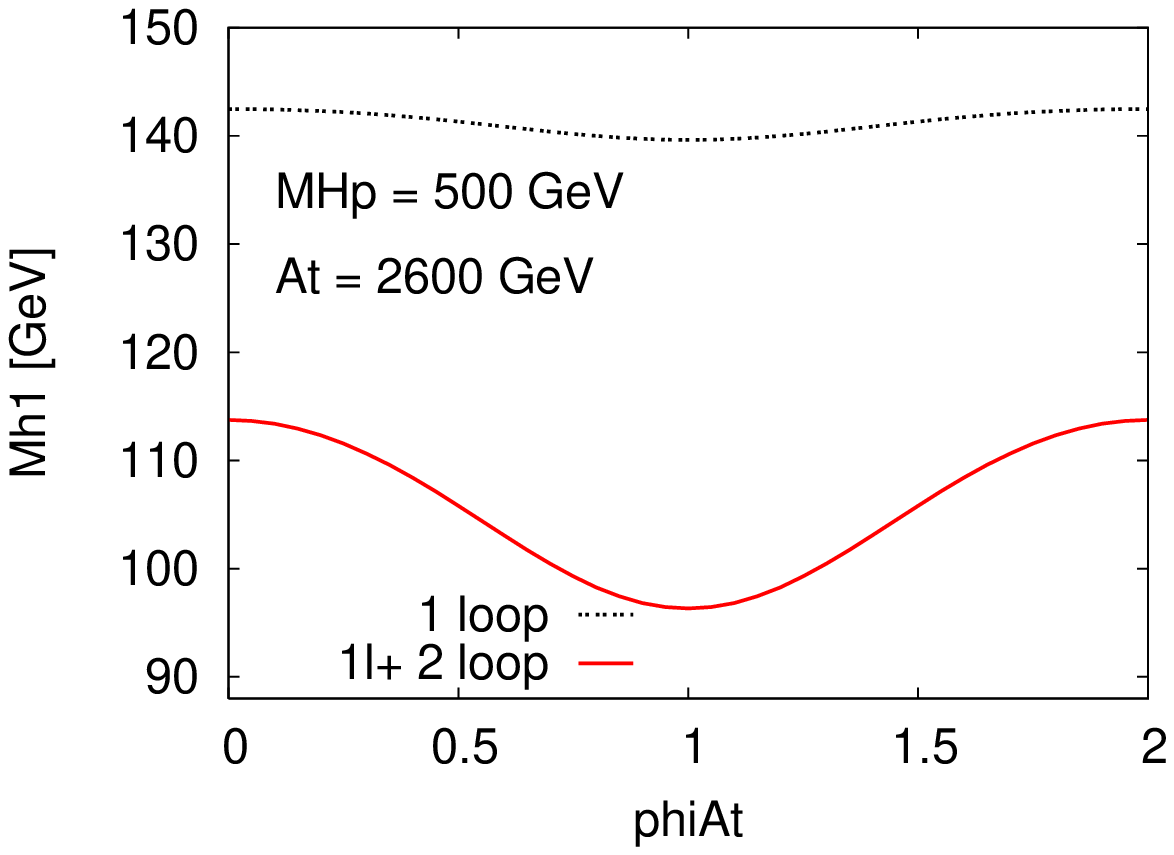}
}
\caption{The lightest Higgs-boson mass, $\MHe$, as a function of
$\phiat$ for $|\At| = 2.6 \tev$ and $\MHp = 500 \gev$. The one-loop 
result (dashed line) is
compared with the result including the \order{\alt\als} corrections 
(solid line). The other parameters are $\msusy = 1000 \gev$,
$\mu = 1000 \gev$, $M_2 = 500 \gev$, 
$\mgl = 1000 \gev$, $\tb = 10$.
}
\label{fig:1}
\end{figure}

The leading two-loop corrections of \order{\alt\als} have been recently
been obtained~\cite{mhcMSSM2L} in the Feynman-diagrammatic approach
for propagator-type corrections, which contribute to the
predictions for the Higgs-boson masses, to wave function normalisation
factors of external Higgs bosons and to effective couplings.
The results are 
valid for arbitrary values of the complex parameters. The impact of the
complex phases of the trilinear coupling $\At$ and the
gluino mass parameter $M_3$ at the two-loop level turns out to be
numerically sizable. As an example, in \reffi{fig:1} the lightest
Higgs-boson mass, $\MHe$, is shown as a function of the phase $\phiat$
of the trilinear coupling $\At$. The one-loop
result (dotted line) is compared with the new result that includes the
\order{\alt\als} contributions (solid line). The dependence on the
complex phase $\phiat$ is much more
pronounced in the two-loop result than in the one-loop case, which can
easily be understood from the analytical structure of the 
corrections~\cite{mhcMSSM2L}. Thus, varying $\phiat$ can give rise to 
shifts in the prediction for 
$\MHe$ of more than $\pm 5 \gev$ even in cases where the 
dependence on the complex phases in 
the one-loop result is very small. The new corrections have
recently been implemented into the 
program {\tt FeynHiggs}~\cite{mhcMSSMlong,mhiggsAEC,feynhiggs}.


\section{Complete one-loop results for Higgs cascade decays}

\begin{figure}[htb!]
\centerline{
\includegraphics[width=0.49\textwidth,height=0.49\textwidth
]{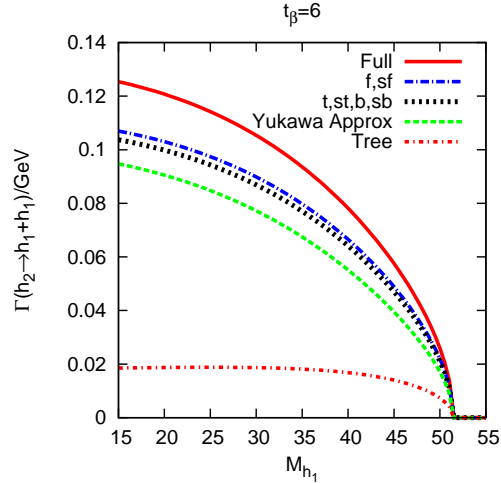}
}
\caption{
The full result for $\Ga(h_2 \to h_1h_1)$ as a function of 
$\MHe$ in the CPX scenario~\cite{cpx} for $\tb = 6$ 
($\MHp$ is varied)
is compared with various
approximations, see text.
}
\label{fig:2}
\end{figure}

For Higgs cascade decays of the kind $h_a \to h_b h_c$, where 
$a,b,c = 1,2,3$, recently complete one-loop results have been obtained in the
cMSSM~\cite{hdeccpv}. They have been supplemented with the
state-of-the-art propagator-type corrections (see above), yielding the
currently most precise prediction for this class
of processes. The genuine vertex corrections turn out to be very
important, yielding a large increase of the decay width compared to a
prediction based on only the tree-level vertex dressed with
propagator-type corrections. This is demonstrated in \reffi{fig:2},
where the full result for $\Ga(h_2 \to h_1h_1)$ as a function of
$\MHe$ in the CPX scenario~\cite{cpx} is compared with results based 
on various approximations for the genuine contributions to the 
$h_2h_1h_1$ vertex. The complete result
(denoted as `Full') differs by more than a factor of six in this example
(for values of $\MHe$ sufficiently below the kinematic limit of
$\MHe = 0.5 \MHz$ where the decay width goes to zero)
from the result for the case where only wave-function
normalisation factors but no genuine one-loop vertex contributions are
taken into account (`Tree').
See \citere{hdeccpv} for a discussion of the other approximations shown
in \reffi{fig:2}.

The new results for the Higgs cascade decays~\cite{hdeccpv}
have been used to analyse the 
impact of the limits on topological cross sections obtained from the
LEP Higgs searches on the parameter space with a very light Higgs boson
within the cMSSM.
It has been found for the
example of the CPX~scenario~\cite{cpx} that, over a large
part of the parameter space where the decay $h_2\to h_1h_1$ is
kinematically possible, it is the dominant decay channel. 
A parameter region
with $\MHe \approx 45 \gev$ and $\tb \approx 6$
remains unexcluded by the limits on topological cross sections
obtained from the LEP Higgs searches, confirming the results of the four
LEP collaborations achieved in a dedicated analysis of the CPX benchmark
scenario. The results of \citere{hdeccpv} will be incorporated into the
public code {\tt FeynHiggs}.


\subsection*{Acknowledgments}

We thank F.~von der Pahlen for useful discussions.
Work supported in part by the European Community's Marie-Curie Research
Training Network under contract MRTN-CT-2006-035505
`Tools and Precision Calculations for Physics Discoveries at Colliders'.




\begin{footnotesize}


\end{footnotesize}

\end{document}